# Branching ratios of $B_c$ Meson Decaying to Pseudoscalar and Axial-Vector Mesons

Neelesh Sharma*, Rohit Dhir and R.C. Verma
Department of Physics, Punjabi University, Patiala-147002, India
*E-mail:nishu.vats@gmail.com.

## Abstract

We study Cabibbo-Kobayashi-Maskawa (CKM) favored weak decays of $B_c$ mesons in the Isgur-Scora-Grinstein-Wise (ISGW) quark model. We present a detailed analysis of the $B_c$ meson decaying to a pseudoscalar meson ($P$) and an axial-vector meson ($A$). We also give the form factors involving $B_c \to A$ transition in the ISGW II framework and consequently, predict the branching ratios of $B_c \to PA$ decays.

## I. INTRODUCTION

The $B_c$ meson discovered at Fermilab [1] is the only quark-antiquark bound system $(\bar{b}c)$ composed of heavy quarks $(b,c)$ with different flavors, and are thus flavor asymmetric. Recently, CDF Collaboration [2] announced an accurate determination of the $B_c$ meson mass and its life time. The investigation of the $B_c$ meson properties (mass spectrum, decay rates, etc.) is therefore of special interest compared to symmetric heavy quarkonium $(\bar{b}b, \bar{c}c)$ states. The decay processes of the $B_c$ meson can be broadly divided into two classes: involving the decay of $b$ quark, $c$ quark, besides the relatively suppressed annihilation of $b$ and $\bar{c}$. Preliminary estimates of the widths of some decay channels of $B_c$ have been made to show that the bound state effects may be significant in $B_c$ decays. Already there exists an extensive literature [3-14] for the semileptonic and nonleptonic decays of the meson $B_c$ involving $s$-wave mesons in the final state. However, relatively less work is done on the $p$-wave emitting weak decays of $B_c$ meson. The decays of $B_c$ meson to a $p$-wave state have been considered previously by other authors [15]. Experimentally, study of the $B_c$ meson decays are in plan for B-Physics both at the TEVATRON and Large Hadron Collider (LHC).

In our recent work [14], we have successfully employed the effects of flavor dependence on $B_c \to P/V$ form factors in BSW model framework and predicted the branching ratios of $B_c \to PP/PV$ decays. In the present work, we employ the Isgur-Scora-Grinstein-Wise (ISGW) II framework [16, 17] to calculate the form factors involving $B_c \to A$ transition. Using the factorization hypothesis, we obtain the decay amplitudes of $B_c$ meson decaying to pseudoscalar and axial-vector mesons and hence, we predict their branching ratios.

The present paper is organized as follows. We give the meson spectroscopy and the methodology in Sections II and III, respectively. We present the brief account of $B_c \to A/P$ transitions form factor and decay constant in Section IV and Section V. Consequently, the branching ratios are estimated. We present the numerical results and discussions in Sec VI and last Section contains summary and conclusions.

## II. MESON SPECTROSCOPY

Both types of axial-vector mesons, ${}^3P_1(J^{PC}=1^{++})$ and ${}^1P_1\,(J^{PC}=1^{+-})$, behave well with respect to the quark model $q\bar{q}$ assignments. Strange and charmed states are most likely a mixture of ${}^3P_1$ and ${}^1P_1$ states, since there is no quantum number forbidding such mixing. In contrast, diagonal ${}^3P_1$ and ${}^1P_1$ systems have opposite C-parity and cannot mix. Experimentally [18], the following non-strange and uncharmed mesons have been observed:

(i) for ${}^3P_1$ multiplet, isovector $a_1(1.230)$ and three isoscalars $f_1(1.285)$, $f_1'(1.512)$ and $\chi_{c1}(3.511)$;

(ii) for ${}^1P_1$ multiplet, isovector $b_1(1.229)$ and three isoscalars $h_1(1.170)$, $h_1'(1.380)$ and $h_{c1}(3.526)$. C-parity of $h_1'(1.380)$ and spin and parity of the $h_{c1}(3.526)$ remains to be confirmed.

Numbers given in the brackets indicate mass (in GeV) of the respective mesons. In the present analysis, mixing of the isoscalar states of ($1^{++}$) mesons is defined as

$$
\begin{aligned}
f_1(1.285) &= \frac{1}{\sqrt{2}}(u\bar{u}+d\bar{d})\cos\phi_A + (s\bar{s})\sin\phi_A, \\
f_1'(1.512) &= \frac{1}{\sqrt{2}}(u\bar{u}+d\bar{d})\sin\phi_A - (s\bar{s})\cos\phi_A, \\
\chi_{c1}(3.511) &= (c\bar{c}),
\end{aligned}
\tag{1}
$$

where

$$\phi_A = \theta(ideal) - \theta_A(physical).$$

Similarly, mixing of two isoscalar meosns $h_1(1.170)$ and $h_1'(1.380)$ is defined as:

$$
\begin{aligned}
h_1(1.170) &= \frac{1}{\sqrt{2}}(u\bar{u}+d\bar{d})\cos\phi_{A'} + (s\bar{s})\sin\phi_{A'}, \\
h_1'(1.380) &= \frac{1}{\sqrt{2}}(u\bar{u}+d\bar{d})\sin\phi_{A'} - (s\bar{s})\cos\phi_{A'}, \\
h_{c1}(3.526) &= (c\bar{c}).
\end{aligned}
\tag{2}
$$

Proximity of $a_1(1.230)$ and $f_1(1.285)$ and to lesser extent that of $b_1(1.229)$ and $h_1(1.170)$ indicates the ideal mixing for both $1^{++}$ and $1^{+-}$ nonets i.e.,

$$\phi_A = \phi_{A'} = 0^\circ. \tag{3}$$

States involving a strange quark of $A(J^{PC}=1^{++})$ and $B(J^{PC}=1^{+-})$ mesons mix to generate the physical states in the following manner:

$$
\begin{aligned}
K_1(1.270) &= K_{1A}\sin\theta_1 + K_{1B}\cos\theta_1, \\
\underline{K}_1(1.400) &= K_{1A}\cos\theta_1 - K_{1B}\sin\theta_1.
\end{aligned}
\tag{4}
$$

where $K_{1A}$ and $K_{1B}$ denote the strange partners of $a_1(1.230)$ and $b_1(1.229)$ respectively. Particle Data Group [18] assumes that the mixing is maximal, i.e., $\theta_1 = 45^0$, whereas $\tau \to K_1(1.270)/K_1(1.400)+\nu_\tau$ data yields $\theta_1 = \pm 37^0$ and $\theta_1 = \pm 58^0$ [19]. However, the study of $D \to K_1(1.270)\pi, K_1(1.400)\pi$ decays rules out positive mixing-angle solutions. Therefore, both negative mixing-angle solutions are allowed by experiment as discussed in detail in [20]. But $D \to K_1^-(1.400)\pi^+$ is largely suppressed for $\theta_1 = -37^0$ and favor the other solution $\theta_1 = -58^0$ [20]. Hence, we take $\theta_1 = -58^0$ in our analysis.

The mixing of charmed and strange charmed states is given by

$$\begin{aligned} D_1(2.422) &= D_{1A} \sin\theta_{D_1} + D_{1B} \cos\theta_{D_1}, \\ \underline{D}_1(2.427) &= D_{1A} \cos\theta_{D_1} - D_{1B} \sin\theta_{D_1}, \end{aligned} \tag{5}$$

and

$$\begin{aligned} D_{s1}(2.460) &= D_{s1A} \sin\theta_{D_{s1}} + D_{s1B} \cos\theta_{D_{s1}}, \\ \underline{D}_{s1}(2.536) &= D_{s1A} \cos\theta_{D_{s1}} - D_{s1B} \sin\theta_{D_{s1}}, \end{aligned} \tag{6}$$

However, in the heavy quark limit, the physical mass eigenstates with $J^P = 1^+$ are $P_1^{3/2}$ and $P_1^{1/2}$ rather than ${}^3P_1$ and ${}^1P_1$ states as the heavy quark spin $S_Q$ decouples from the other degrees of freedom so that $S_Q$ and the total angular momentum of the light antiquark are separately good quantum numbers. Thererfore, we can write

$$\begin{aligned} |P_1^{1/2}> &= -\sqrt{\frac{1}{3}}\,|\,{}^1P_1> + \sqrt{\frac{2}{3}}\,|\,{}^3P_1>, \\ |P_1^{3/2}> &= \sqrt{\frac{2}{3}}\,|\,{}^1P_1> + \sqrt{\frac{1}{3}}\,|\,{}^3P_1>. \end{aligned} \tag{7}$$

.

Hence, the states $D_1(2.422)$ and $\underline{D}_1(2.427)$ can be identified with $P_1^{1/2}$ and $P_1^{3/2}$, respectively. However, beyond the heavy quark limit, there is a mixing between $P_1^{1/2}$ and $P_1^{3/2}$, denoted by

$$\begin{aligned} D_1(2.422) &= D_1^{1/2} \cos\theta_2 + D_1^{3/2} \sin\theta_2, \\ \underline{D}_1(2.427) &= -D_1^{1/2} \sin\theta_2 + D_1^{3/2} \cos\theta_2. \end{aligned} \tag{8}$$

Likewise for strange axial-vector charmed mesons,

$$\begin{aligned} D_{s1}(2.460) &= D_{s1}^{1/2} \cos\theta_3 + D_{s1}^{3/2} \sin\theta_3, \\ \underline{D}_{s1}(2.536) &= -D_{s1}^{1/2} \sin\theta_3 + D_{s1}^{3/2} \cos\theta_3. \end{aligned} \tag{9}$$

The mixing angle $\theta_2 = (5.7 \pm 2.4)^0$ is obtained by Belle through a detailed $B \to D^*\pi\pi$ analysis [21], while $\theta_3 \approx 7^0$ is determined from the quark potential model [22, 23].

Therefore, for $b$-flavored mesons we use the predictions of quark model analysis. Since the quark model analysis has been quite successful in explaining the mixing of strange and charmed states. Therefore in this case also we use the following:

$$\begin{aligned} B_1(5.670) &= B_{1A}\sin\theta_4 + B_{1A'}\cos\theta_4, \\ \underline{B}_1(5.721) &= B_{1A}\cos\theta_4 - B_{1A'}\sin\theta_4, \end{aligned} \tag{10}$$

and

$$\begin{aligned} B_{s1}(5.762) &= B_{s1A}\sin\theta_5 + B_{s1A'}\cos\theta_5, \\ \underline{B}_{s1}(5.830) &= B_{s1A}\cos\theta_5 - B_{s1A'}\sin\theta_5, \end{aligned} \tag{11}$$

For bottom states, we have taken masses from the review of particle properties [18]. We use the following mixing angles $\theta_4 \approx -43^0$ and $\theta_5 \approx -45^0$.

For $\eta$ and $\eta'$ pseudoscalar states, we use

$$\begin{aligned} \eta(0.547) &= \frac{1}{\sqrt{2}}(u\bar{u}+d\bar{d})\sin\phi_P - (s\bar{s})\cos\phi_P, \\ \eta'(0.958) &= \frac{1}{\sqrt{2}}(u\bar{u}+d\bar{d})\cos\phi_P + (s\bar{s})\sin\phi_P, \end{aligned} \tag{12}$$

where $\phi_P = \theta(ideal) - \theta_P(physical)$ and $\theta_P(physical) = -15.4^\circ$.

## III. METHODOLOGY

### A. Weak Hamiltonian

To the lowest order in weak interaction, the non-leptonic Hamiltonian has the usual current ⊗ current form

$$H_w = \frac{G_F}{\sqrt{2}} J^{+}{}_{\mu} J^{\mu} + h.c. \tag{13}$$

The weak current $J_\mu$ is given by

$$J_\mu = (\bar{u}\ \bar{c}\ \bar{t})\,\gamma_\mu(1\text{-}\gamma_5)\begin{pmatrix} d' \\ s' \\ b' \end{pmatrix}, \tag{14}$$

where $d'$, $s'$, and $b'$ are mixture of the $d, s,$ and $b$ quarks, as given by Cabibbo-Kobayashi-Maskawa (CKM) matrix [18].

**a) For bottom changing decays**

The QCD modified weak Hamiltonian [24] generating the *b*-quark decays in CKM enhanced modes ($\Delta b = 1$, $\Delta C = 1$, $\Delta S = 0$; $\Delta b = 1$, $\Delta C = 0$, $\Delta S = -1$) is given by

$$H_w^{\Delta b=1} = \frac{G_F}{\sqrt{2}}\{V_{cb}V_{ud}^*[c_1(\mu)(\bar{c}b)(\bar{d}u) + c_2(\mu)(\bar{c}u)(\bar{d}b)] + V_{cb}V_{cs}^*[c_1(\mu)(\bar{c}b)(\bar{s}c) + c_2(\mu)(\bar{c}c)(\bar{s}b)]\}, \tag{15}$$

where $G_F$ is the Fermi constant and $V_{ij}$ are the CKM matrix elements, $c_1$ and $c_2$ are the standard perturbative QCD coefficients.

**b) For bottom conserving and charm changing decays**

In addition to the bottom changing decays, bottom conserving decay channel is also available for the $B_c$ meson, where the charm quark decays to an *s* or *d* quark. The weak Hamiltonian generating the *c*-quark decays in CKM enhanced mode ($\Delta b = 0$, $\Delta C = -1$, $\Delta S = -1$) is given by

$$H_w^{\Delta c=-1} = \frac{G_F}{\sqrt{2}}V_{ud}V_{cs}^*\,[c_1(\mu)(\bar{u}d)(\bar{s}c) + c_2(\mu)(\bar{u}c)(\bar{s}d)]. \tag{16}$$

One naively expects this channel to be suppressed kinematically due to the small phase space available. However, the kinematic suppression is well compensated by the CKM element $V_{cs}$, which is larger than $V_{cb}$ appearing for the bottom changing decays. In fact, we shall show later that bottom conserving decay modes are more prominent than the bottom changing ones.

By factorizing matrix elements of the four-quark operator contained in the effective Hamiltonian (12) and (13), one can distinguish three classes of decays [25]:

- The first class contains those decays which can be generated from color singlet current and the decay amplitudes are proportional to $a_1$, where $a_1(\mu) = c_1(\mu) + \frac{1}{N_c}c_2(\mu)$, and $N_c$ is the number of colors.
- A second class of transitions consist of those decays which can be generated from neutral current. The decay amplitude in this class is proportional to $a_2$ i.e. for the color suppressed modes $a_2(\mu) = c_2(\mu) + \frac{1}{N_c}c_1(\mu)$.
- The third class of decay modes can be generated from the interference of color singlet and color neutral currents i.e. the $a_1$ and $a_2$ amplitudes interfere.

However, we follow the convention of large $N_c$ limit to fix QCD coefficients $a_1 \approx c_1$ and $a_2 \approx c_2$, where [24]:

$$c_1(\mu) = 1.26 \ , \ c_2(\mu) = -0.51 \ \text{at} \ \mu \approx m_c^2 ,$$
$$c_1(\mu) = 1.12 \ , \ c_2(\mu) = -0.26 \ \text{at} \ \mu \approx m_b^2 . \tag{17}$$

**B. Decay Amplitudes and Rates**

The decay rate formula for $B_c \to PA$ decays is given by

$$\Gamma(B_c \to P\,A) = \frac{p_c^3}{8\pi\, m_A^2} \left| A(B_c \to P\,A) \right|^2 , \tag{18}$$

where $p_c$ is the magnitude of the three-momentum of a final-state particle in the rest frame of $B_c$ meson and $m_A$ denotes the mass of the axial-vector meson.

In the naive factorization hypothesis the decay amplitudes can be expressed as a product of the matrix elements of weak currents (up to the weak scale factor of $\frac{G_F}{\sqrt{2}}$ $\times$ CKM elements$\times$QCD factor) given by

$$\begin{aligned} \langle PA | H_w | B_c \rangle &\sim \langle P | J^\mu | 0 \rangle \langle A | J_\mu | B_c \rangle + \langle A | J^\mu | 0 \rangle \langle P | J_\mu | B_c \rangle , \\ \langle PA' | H_w | B_c \rangle &\sim \langle P | J^\mu | 0 \rangle \langle A' | J_\mu | B_c \rangle + \langle A' | J^\mu | 0 \rangle \langle P | J_\mu | B_c \rangle . \end{aligned} \tag{19}$$

Using Lorentz invariance, matrix elements of the current between meson states can be expressed [16, 17, 25] as

$$\begin{aligned} \langle P | J_\mu | 0 \rangle &= -i f_P k_\mu , \\ \langle A | J_\mu | 0 \rangle &= \in_\mu^* m_A f_A , \\ \langle A' | J_\mu | 0 \rangle &= \in_\mu^* m_{A'} f_{A'} , \\ \langle A(k_A) | J_\mu | B_c(k_{B_c}) \rangle &= l \in_\mu^* + c_+ (\in^* \cdot k_{B_c})(k_{B_c} + k_A)_\mu + c_- (\in^* \cdot k_{B_c})(k_{B_c} - k_A)_\mu , \\ \langle A'(k_{A'}) | J_\mu | B_c(k_{B_c}) \rangle &= r \in_\mu^* + s_+ (\in^* \cdot k_{B_c})(k_{B_c} + k_{A'})_\mu + s_- (\in^* \cdot k_{B_c})(k_{B_c} - k_{A'})_\mu , \end{aligned} \tag{20}$$

and

$$\langle P(k_P) | J_\mu | B_c(k_{B_c}) \rangle = (k_{B_c \mu} + k_{P\mu} - \frac{m_{B_c}^2 - m_P^2}{q^2} q_\mu)\, F_1^{B_c P}(q^2) + \frac{m_{B_c}^2 - m_P^2}{q^2} q_\mu\, F_0^{B_c P}(q^2).$$

Which yield

$$\begin{aligned} A(B_c \to PA) &= (\, 2\, m_A \ f_A\, F_1^{B_c \to P}(m_A^2) \ + \ f_P \ F^{B_c \to A}(m_P^2)\,) , \\ A(B_c \to PA') &= (\, 2\, m_{A'} \ f_{A'}\, F_1^{B_c \to P}(m_{A'}^2) \ + \ f_P \ F^{B_c \to A'}(m_P^2)\,) , \end{aligned} \tag{21}$$

where

$$F^{B_c\to A}(m_P^2)=l+(m_{B_c}^2-m_A^2)\,c_+ + m_P^2\,c_-,$$
$$F^{B_c\to A'}(m_P^2)=r+(m_{B_c}^2-m_{A'}^2)\,s_+ + m_P^2\,s_-. \qquad (22)$$

Sandwiching the weak Hamiltonian (15) and (16) between the initial and the final states, the decay amplitudes for various $B_c \to PA$ decay modes are obtained, which are given in Tables I and II for the following three categories:

I) involving $P(0^-)\to P(0^-)$ transitions only,

II) involving $P(0^-)\to A(1^+)$ transitions only, and

III) involving both $P(0^-)\to P(0^-)/A(1^+)$ transitions.

## IV. FORM FACTORS

In this section, we give a short description of ISGW II model [16], in order to calculate the $B_c \to A$ transition form factors. Consider the transition $B_c \to A$, where the axial vector meson [16, 17] has the quark content $q_1\bar{q}_2$ with $\bar{q}_2$ being the spectator quark. We begin with the definition

$$F_n=\left(\frac{\tilde{m}_A}{\tilde{m}_{B_c}}\right)^{1/2}\left(\frac{\beta_{B_c}\beta_A}{B_{B_cA}}\right)^n\left[1+\frac{1}{18}h^2(t_m-t)\right]^{-3}, \qquad (23)$$

where

$$h^2=\frac{3}{4m_cm_1}+\frac{3m_2^2}{2\overline{m}_{B_c}\overline{m}_A\beta_{B_cA}^2}+\frac{1}{\overline{m}_{B_c}\overline{m}_A}(\frac{16}{33-2n_f})\ln[\frac{\alpha_S(\mu_{QM})}{\alpha_S(m_1)}] \qquad (24)$$

$\tilde{m}$ is the sum of the mesons constituent quarks masses, $\overline{m}$ is the averaged mass, $t_m=(m_{B_c}-m_A)^2$ is the maximum momentum transfer and

$$\mu_\pm=(\frac{1}{m_1}+\frac{1}{m_b})^{-1}. \qquad (25)$$

with $m_1$ and $m_2$ being the masses of the quark $q_1$ and $\bar{q}_2$ respectively. In (23), the values of the parameters $\beta_{B_c}$ and $\beta_A$ related as $\beta_{B_cA}^2=1/2(\beta_{B_c}^2+\beta_A^2)$, for more details see [16].

The form factors have the following expressions in the ISGW II model [16].

$$l=\tilde{m}_{B_c}\beta_{B_c}[\frac{1}{\mu_-}+\frac{m_2\tilde{m}_A(\tilde{\omega}-1)}{\beta_{B_c}^2}(\frac{5+\tilde{\omega}}{6m_1}-\frac{m_2\beta_{B_c}^2}{2\mu_-\beta_{B_cA}^2})]F_5^{(l)},$$

$$c_+ + c_- = -\frac{m_2\tilde{m}_A}{2m_1m_2\beta_{B_c}}\left(1-\frac{m_1m_2\beta^2_{B_c}}{2m_A\mu_-\beta^2_{B_cA}}\right)F^{(c_++c_-)},$$

$$c_+ - c_- = -\frac{m_2\tilde{m}_A}{2m_1m_2\beta_{B_c}}\left(\frac{\tilde{\omega}+2}{3}-\frac{m_1m_2\beta^2_{B_c}}{2m_A\mu_-\beta^2_{B_cA}}\right)F^{(c_+-c_-)}, \tag{26}$$

where

$$F_5^{(l)} = F_5(\frac{\bar{m}_{B_c}}{\tilde{m}_{B_c}})^{1/2}(\frac{\bar{m}_A}{\tilde{m}_A})^{1/2},$$

$$F_5^{(c_++c_-)} = F_5(\frac{\bar{m}_{B_c}}{\tilde{m}_{B_c}})^{-3/2}(\frac{\bar{m}_A}{\tilde{m}_A})^{1/2}, \tag{27}$$

$$F_5^{(c_+-c_-)} = F_5(\frac{\bar{m}_{B_c}}{\tilde{m}_{B_c}})^{-1/2}(\frac{\bar{m}_A}{\tilde{m}_A})^{-1/2},$$

and

$$\tilde{\omega}-1=\frac{t_m-t}{2\bar{m}_{B_c}\bar{m}_A}. \tag{28}$$

In the original version of the ISGW I model [16], the function $F_n$ has a different expression in its $t_m - t$ dependence.

$$F_n = \left(\frac{\tilde{m}_A}{\tilde{m}_{B_c}}\right)^{1/2}\left(\frac{\beta_{B_c}\beta_A}{\beta_{B_cA}}\right)^n \exp\{-\frac{m_2}{4\tilde{m}_{B_c}\tilde{m}_A}\frac{(t_m-t)}{k^2\beta^2_{B_cA}}\}, \tag{29}$$

where $k = 0.7$ the relativistic correction factor. The form factors are given by

$$l = -\tilde{m}_{B_c}\beta_{B_c}[\frac{1}{\mu_-}+\frac{m_2(t_m-t)}{2\tilde{m}_{B_c}k^2\beta^2_{B_c}}(\frac{1}{m_1}-\frac{m_2\beta^2_{B_c}}{2\mu_-\tilde{m}_A\beta^2_{B_cA}})]F_5,$$

$$c_+ = -\frac{m_2m_c}{4\tilde{m}_{B_c}\mu_-\beta_{B_c}}(1-\frac{m_1m_2\beta^2_{B_c}}{2\mu_-\tilde{m}_A\beta^2_{B_cA}})]F_5. \tag{30}$$

It is clear that the form factor $c_+$ has an opposite sign in the ISGW I and ISGW II models. Note that, in the ISGW2 model [16] allows one to determine the form factor factors $c_-$ and $s_-$, which vanish in the ISGW I model. We use the following quark masses

$$m_u = m_d = 0.31,\ m_s = 0.49,\ m_c = 1.7, \text{ and } m_b = 5.0,$$

obtained from meson masses, to calculate the form factors for $B_c \to A$ and $B_c \to A'$ transitions. The obtained form factors are given in Tables III and IV.

For $B_c \to P$ transition, we use the form factors obtained in our earlier work [14], where we have employed the effects of flavor dependence on these form factors using BSW model framework. For the sake of convenience, we have given these form factors in column 3 of Table V.

## V. DECAY CONSTANTS

Decay constant of pseudoscalar mesons are well known. In this work, we use the following values of decay constants [5, 18, and 26] of the pseudoscalar mesons $(0^-)$ :

$$\begin{aligned} &f_\pi = 0.131 \text{ GeV},\ f_K = 0.160 \text{ GeV},\ f_D = 0.208 \text{ GeV}, \\ &f_{D_s} = 0.273 \text{ GeV},\quad f_\eta = 0.133\ GeV, \\ &\text{and } f_{\eta'} = 0.126\ GeV. \end{aligned} \tag{31}$$

However, for axial-vector meson, decay constants for $J^{PC} = 1^{+-}$ mesons may vanish due to the C-parity behavior. Under charge conjunction, the two types of axial-vector mesons transform as

$$\begin{aligned} M_b^a\ (1^{++}) &\to +M_a^b(1^{++}) \\ M_b^a\ (1^{+-}) &\to -M_a^b(1^{+-}) \end{aligned} \qquad (a, b = 1, 2, 3)$$

Where $M_b^a$ denotes meson 3×3 matrix elements in SU(3) flavor symmetry. Since the weak axial-vector current transforms as $(A_\mu)_b^a \to +(A_\mu)_a^b$ under charge conjunction, only the $(1^{++})$ state can be produced through the axial-vector current in the SU(3) symmetry limit [24]. Particle Data Group [18] assumes that the mixing is maximal, i.e., $\theta = 45^0$, whereas $\tau \to K_1(1.270)/K_1(1.400) + \nu_\tau$ data yields $\theta = \pm 37^0$ and $\theta = \pm 58^0$ [22, 24]. To determine the decay constant of $K_1(1.270)$, we use the following formula:

$$\Gamma(\tau \to K_1 \nu_\tau) = \frac{G_F^2}{16\pi} |V_{us}|^2\, f_{K_1}^2\, \frac{(m_\tau^2 + 2m_{K_1}^2)(m_\tau^2 - m_{K_1}^2)^2}{m_\tau^3}, \tag{32}$$

which gives $f_{K_1(1270)} = 0.175 \pm 0.019$ GeV. The decay constant of $K_1(1.400)$ can be obtained from $f_{K_1(1.400)} / f_{K_1(1.270)} = \cot\theta$. A small value around 0.011 GeV for the decay constant of $K_{1B}$ may arise through SU(3) breaking, which yields

$$\begin{aligned} f_{\underline{K}_1(1.400)} &= f_{K_{1A}} \cos\theta_1 - f_{K_{1B}} \sin\theta_1 \\ &= -0.087\, GeV, \end{aligned} \tag{33}$$

for $\theta_1 = -58^0$ [22]. Similarly, decay constant of $a_1(1.260)$ can be obtained from $B(\tau \to a_1\nu_\tau)$. However, this branching ratio is not given in Particle Data Group [18], although the data on $\tau \to a_1\nu_\tau \to \rho\pi\nu_\tau$ have been reported by various experiments. We take $f_{a_1} = 0.203 \pm 0.018$ GeV from the analysis given by J.C.R. Bloch *et. al* [27]. For the decay constant of $f_1(1.285)$, we assume $f_{f_1} \approx f_{a_1}$. The decay constants

$$f_{D_{1A}} = -0.127\ GeV,\ f_{D_{1B}} = 0.045\ GeV,$$
$$f_{D_{s1A}} = -0.121\ GeV,\ f_{D_{s1B}} = 0.038\ GeV, \qquad (34)$$

have been taken from [22] and determine $f_{\chi_{c1}} \approx -0.160\ GeV$ [22].

## VI. NUMERICAL RESULTS INCLUDING THE FLAVOR DEPENDENCE

Using $B_c \to A/P$ form factors obtained in section IV, we finally predict branching ratios of various $B_c \to PA$ decays as shown in Tables VI and VII. We observe the following:

i) Naively, the *c* → *u, s* (charm changing and bottom conserving) decay channels are expected to be kinematically suppressed, however, the large value of the CKM matrix elements along with the large value of *c* → *u, s* transition form factors overcome this suppression. As a result, branching ratios of the charm changing mode are enhanced as compare to the bottom changing modes.

ii) The dominant decay for charm changing and bottom conserving transitions in $(\Delta b = 0, \Delta C = -1, \Delta S = -1)$ are: $B(B_c^+ \to \pi^+ B_{s1}^0)$ = 3.57%, $B(B_c^+ \to \bar{K}^0 B_1^+)$ = 0.62% and $B(B_c^+ \to \pi^+ \underline{B}_{s1}^0)$ = 0.21%. The next order dominant decay is $B$ $(B_c^+ \to \bar{K}^0 \underline{B}_1^+)$ = 0.05%.

iii) For bottom changing transitions the dominating decays in $(\Delta b = 1, \Delta C = 1, \Delta S = 0)$ mode is $B(B_c^- \to \eta_c a_1^-)$ = 0.28%, and in $(\Delta b = 1, \Delta C = 0, \Delta S = -1)$ mode are $B(B_c^- \to D_s^- h_{c1})$ = 0.15% and $B(B_c^- \to D_s^- \chi_{c1})$ = 0.12%. The rest of the decay modes (see Table V) remain suppressed partly due to the small values of the CKM matrix elements and the small values of the form factors.

iv) In contrast to the charm meson sector, the experimental data of *B* meson decays favor the constructive interference between color favored and color suppressed diagrams [20], giving $a_1 = 1.10 \pm 0.08$ and $a_2 = 0.20 \pm 0.02$. Taking $a_1 = 1.10$ and $a_2 = 0.20$ for the constructive interference case, we obtain larger value for $B(B_c^- \to D_s^- \chi_{c1})$ = 0.14% in comparison to 0.12% (for destructive interference).

## VII. CONCLUSIONS

In this paper, we have studied hadronic weak decays of $B_c \to PA$ in Cabibbo-favored channel. At present, no experimental information is available on these decay modes. We make the following conclusions:

i) One naively expects the bottom conserving modes to be kinematically suppressed; however, relatively the large value of the CKM matrix elements accompanied with large values of the form factors compensates the suppression.

ii) Because of the small values of the form factors the decay modes involving the $b \to u, c$ transitions are suppressed in comparison to $c \to u, c$ decay channels. However, the small values of the CKM matrix elements further suppress these decays.

iii) The dominant decays for charm changing and bottom conserving mode are: $B(B_c^+ \to \pi^+ B_{s1}^0)$ , $B(B_c^+ \to \bar{K}^0 B_1^+)$ and $B(B_c^+ \to \pi^+ \underline{B}_{s1}^0)$. For bottom changing modes the dominant decays are: $B(B_c^- \to \eta_c a_1^-)$, $B(B_c^- \to D_s^- h_{c1})$ and $B(B_c^- \to D_s^- \chi_{c1})$ .

iv) Taking in to account the constructive interference between color favored and color suppressed diagrams, we observe that the branching ratio of $B_c^- \to D_s^- \chi_{c1}$ decay gets further enhanced to 0.14%.

v) We hope our prediction for branching ratios are useful from experimental point of view. Observation of these $B_c$ processes in the $B_c$ experiments such as Belle, Babar, BTeV, LHC and so on will be crucial in testing the ISGW quark model as well as validity of the factorization scheme. Measurements of these branching ratios present an interesting test of our predictions.

## ACKNOWLEDGEMENT

One of the authors (N.S.) is thankful to the University Grant Commission, New Delhi, for the financial assistance.

**Table I. Decay amplitudes of CKM-favored mode of $B_c \to PA$ decays for Charm changing modes.**

| Decays | Amplitude |
|---|---|
| | $\Delta b$ =0, $\Delta C$ = -1, $\Delta S$ = -1 |
| **a)** $P(0^-) \to A(1^+)$ **transition** | |
| $B_c^+ \to \pi^+ B_{s1}^0$ | $a_1 f_\pi \sin\theta\, F^{B_c \to B_{s1A}}(m_\pi^2) + a_1 f_\pi \cos\theta\, F^{B_c \to B_{s1A'}}(m_\pi^2)$ |
| $B_c^+ \to \pi^+ \underline{B}_{s1}^0$ | $a_1 f_\pi \cos\theta\, F^{B_c \to B_{s1A}}(m_\pi^2) - a_1 f_\pi \sin\theta\, F^{B_c \to B_{s1A'}}(m_\pi^2)$ |
| $B_c^+ \to \overline{K}^0 B_1^+$ | $a_2 f_K \sin\theta\, F^{B_c \to B_{1A}}(m_\pi^2) + a_2 f_K \cos\theta\, F^{B_c \to B_{1A'}}(m_\pi^2)$ |
| $B_c^+ \to \overline{K}^0 \underline{B}_1^+$ | $a_2 f_K \cos\theta\, F^{B_c \to B_{1A}}(m_\pi^2) + a_2 f_K \sin\theta\, F^{B_c \to B_{1A'}}(m_\pi^2)$ |

**Table II. Decay amplitudes of CKM-favored modes of $B_c \to PA$ decays for Bottom changing modes.**

| Decays | Amplitude |
|---|---|
| | $\Delta b = 1, \Delta C = 1, \Delta S = 0$ |
| **a)** $P(0^-) \to P(0^-)$ **transition** | |
| $B_c^- \to D^- D_1^0$ | $2a_2 m_{D_1} f_{D_{1A}} \sin\theta\, F^{B_c \to D}(m_{D_1}^2) + 2a_2 m_{D_1} f_{D_{1A'}} \cos\theta\, F^{B_c \to D}(m_{D_1}^2)$ |
| $B_c^- \to D^- \underline{D}_1^0$ | $2a_2 m_{\underline{D}_1} f_{D_{1A}} \sin\theta\, F^{B_c \to D}(m_{\underline{D}_1}^2) + 2a_2 m_{\underline{D}_1} f_{D_{1A'}} \cos\theta\, F^{B_c \to D}(m_{\underline{D}_1}^2)$ |
| $B_c^- \to \eta_c a_1^-$ | $2a_1 m_{a_1} f_{a_1} F^{B_c \to \eta_c}(m_{a_1}^2)$ |
| $B_c^- \to \eta_c b_1^-$ | $2a_1 m_{b_1} f_{b_1} F^{B_c \to \eta_c}(m_{b_1}^2)$ |
| **b)** $P(0^-) \to A(1^+)$ **transition** | |
| $B_c^- \to \pi^- \chi_{c1}$ | $a_1 f_\pi F^{B_c \to \chi_{c1}}(m_\pi^2)$ |
| $B_c^- \to \pi^- h_{c1}$ | $a_1 f_\pi F^{B_c \to h_{c1}}(m_\pi^2)$ |
| $B_c^- \to D^0 D_1^-$ | $a_2 f_D \sin\theta\, F^{B_c \to D_{1A}}(m_D^2) + a_2 f_D \cos\theta\, F^{B_c \to D_{1A'}}(m_D^2)$ |
| $B_c^- \to D^0 \underline{D}_1^-$ | $a_2 f_D \cos\theta\, F^{B_c \to D_{1A}}(m_D^2) - a_2 f_D \sin\theta\, F^{B_c \to D_{1A'}}(m_D^2)$ |
| | $\Delta b = 1, \Delta C = 0, \Delta S = -1$ |
| **a)** $P(0^-) \to P(0^-)$ **transition** | |
| $B_c^- \to \overline{D}^0 K_1^-$ | $2a_1 m_{K_1} f_{K_{1A}} \sin\theta\, F^{B_c \to D}(m_{K_1}^2) + 2a_1 m_{K_1} f_{K_{1A'}} \cos\theta\, F^{B_c \to D}(m_{K_1}^2)$ |
| $B_c^- \to \overline{D}^0 \underline{K}_1^-$ | $2a_1 m_{\underline{K}_1} f_{K_{1A}} \cos\theta\, F^{B_c \to D}(m_{\underline{K}_1}^2) - 2a_1 m_{\underline{K}_1} f_{K_{1A'}} \sin\theta\, F^{B_c \to D}(m_{\underline{K}_1}^2)$ |
| $B_c^- \to D_s^- a_1^0$ | $\sqrt{2} a_2 m_{a_1} f_{a_1} F^{B_c \to D_s}(m_{a_1}^2)$ |
| $B_c^- \to D_s^- f_1$ | $\sqrt{2} a_2 m_{f_1} f_{f_1} \cos\theta\, F^{B_c \to D_s}(m_{f_1}^2)$ |
| **b)** $P(0^-) \to A(1^+)$ **transition** | |
| $B_c^- \to K^- \overline{D}_1^0$ | $a_1 m_K f_K \sin\theta\, F^{B_c \to D_{1A}}(m_K^2) + a_1 m_K f_K \cos\theta\, F^{B_c \to D_{1A'}}(m_K^2)$ |
| $B_c^- \to K^- \underline{D}_1^0$ | $a_1 m_K f_K \cos\theta\, F^{B_c \to D_{1A}}(m_K^2) - a_1 m_K f_K \sin\theta\, F^{B_c \to D_{1A'}}(m_K^2)$ |
| $B_c^- \to D_s^- h_{c1}$ | $a_1 f_{D_s} F^{B_c \to h_{c1}}(m_{D_s}^2)$ |
| **c)** $P(0^-) \to P(0^-)/A(1^+)$ **transition** | |
| $B_c^- \to D_s^- \chi_{c1}$ | $2a_2 m_{\chi_{c1}} f_{\chi_{c1}} F^{B_c \to D_s}(m_{\chi_{c1}}^2) + a_1 f_{D_s} F^{B_c \to \chi_{c1}}(m_{D_s}^2)$ |

**Table III. Form factors of $B_c \to A$ transition at $q^2 = t_m$ in the ISGW II quark model**

| Modes | Transition | *l* | $c_+$ | $c_-$ |
|---|---|---|---|---|
| $\Delta b = 0, \Delta C = -1, \Delta S = -1$ | $B_c \to B_{s1}$ | -3.911 | -0.948 | -0.044 |
| | $B_c \to B_1$ | -5.365 | -0.770 | -0.042 |
| $\Delta b = 1, \Delta C = 0, \Delta S = -1$ | $B \to D_1$ | -3.605 | -0.047 | -0.006 |
| | $B_c \to D_{s1}$ | -2.888 | -0.060 | -0.006 |
| $\Delta b = 1, \Delta C = 1, \Delta S = 0$ | $B_c \to \chi_c(c\bar{c})$ | -1.193 | -0.101 | -0.005 |

**Table IV. Form factors of $B_c \to A'$ transition at $q^2 = t_m$ in the ISGW II quark model**

| Modes | Transition | *r* | $s_+$ | $s_-$ |
|---|---|---|---|---|
| $\Delta b = 0, \Delta C = -1, \Delta S = -1$ | $B_c \to B'_{s1}$ | 5.045 | 0.273 | 0.213 |
| | $B_c \to B'_1$ | 5.502 | 0.208 | 0.107 |
| $\Delta b = 1, \Delta C = 0, \Delta S = -1$ | $B \to D'_1$ | 2.882 | 0.082 | - 0.056 |
| | $B_c \to D'_{s1}$ | 2.488 | 0.101 | -0.060 |
| $\Delta b = 1, \Delta C = 1, \Delta S = 0$ | $B_c \to h_{c1}(c\bar{c})$ | 1.690 | 0.141 | -0.040 |

**Tables V. Form factors of $B_c \to P$ transition**

| Modes | Transition | $F_0^{B_cP}(0)$ (using flavor dependent $\omega$) |
|---|---|---|
| $\Delta b = 0, \Delta C = -1, \Delta S = -1$ | $B_c \to B_s$ | 0.55 |
| | $B_c \to B$ | 0.41 |
| $\Delta b = 1, \Delta C = 0, \Delta S = -1$ | $B_c \to D$ | 0.08 |
| | $B_c^- \to D_s$ | 0.15 |
| $\Delta b = 1, \Delta C = 1, \Delta S = 0$ | $B_c \to \eta_c(c\bar{c})$ | 0.58 |

**Table VI. Branching ratios of CKM-favored mode of $B_c \to PA$ decays for Charm changing modes.**

| Decays | Br (%) |
|---|---|
| $\Delta b$ =0, $\Delta C$ = -1, $\Delta S$ = -1 | |
| **a)** $P(0^-) \to A(1^+)$ **transition** | |
| $B_c^+ \to \pi^+ B_{s1}^0$ | 3.57 |
| $B_c^+ \to \pi^+ \underline{B}_{s1}^0$ | 0.21 |
| $B_c^+ \to \bar{K}^0 B_1^+$ | 0.62 |
| $B_c^+ \to \bar{K}^0 \underline{B}_1^+$ | 0.05 |

**Table VII. Branching ratios of CKM-favored modes of $B_c \to PA$ decays for Bottom changing modes.**

| Decays | Br (%) |
|---|---|
| $\Delta b = 1, \Delta C = 1, \Delta S = 0$ | |
| **a)** $P(0^-) \to P(0^-)$ **transition** | |
| $B_c^- \to D^- D_1^0$ | 1.40×10^-4 |
| $B_c^- \to D^- \underline{D}_1^0$ | 3.71×10^-6 |
| $B_c^- \to \eta_c a_1^-$ | 0.28 |
| $B_c^- \to \eta_c b_1^-$ | 2.46×10^-6 |
| **b)** $P(0^-) \to A(1^+)$ **transition** | |
| $B_c^- \to \pi^- \chi_{c1}$ | 0.07 |
| $B_c^- \to \pi^- h_{c1}$ | 0.06 |
| $B_c^- \to D^0 D_1^-$ | 7.87×10^-3 |
| $B_c^- \to D^0 \underline{D}_1^-$ | 5.77×10^-6 |
| $\Delta b = 1, \Delta C = 0, \Delta S = -1$ | |
| **a)** $P(0^-) \to P(0^-)$ **transition** | |
| $B_c^- \to \overline{D}^0 K_1^-$ | 2.67×10^-6 |
| $B_c^- \to \overline{D}^0 \underline{K}_1^-$ | 8.07×10^-7 |
| $B_c^- \to D_s^- a_1^0$ | 3.55×10^-7 |
| $B_c^- \to D_s^- f_1$ | 4.19×10^-7 |
| **b)** $P(0^-) \to A(1^+)$ **transition** | |
| $B_c^- \to K^- \overline{D}_1^0$ | 3.36×10^-5 |
| $B_c^- \to K^- \underline{D}_1^0$ | 1.15×10^-8 |
| $B_c^- \to D_s^- h_{c1}$ | 0.15 |
| **c)** $P(0^-) \to P(0^-) / A(1^+)$ **transition** | |
| $B_c^- \to D_s^- \chi_{c1}$ | 0.12 |